\documentstyle[amsfonts,12pt]{article}
\def\one{1\hskip-.37em 1}

\def\half{\textstyle{\frac{1}{2}}}

\def\D{{\cal D}}

\def\ra{\rightarrow}
\def\tint{{\textstyle\int}}

\def\s{\hskip.1em}

\def\b{\begin{eqnarray*}}     %takes no eqn numbers
\def\e{\end{eqnarray*}}       %takes no eqn numbers
\def\bn{\begin{eqnarray}}     %takes eqn numbers 
\def\en{\end{eqnarray}}       %takes eqn numbers
\def\<{\langle}
\def\>{\rangle}

\def\no{\nonumber}

\def\{{\lbrace}
\def\}{\rbrace}
\begin{document}
%\footnote{Electronic mail: klauder@phys.ufl.edu}
\title{Metric and Curvature in\\ Gravitational Phase Space}
\author{Glenn Watson and John R. Klauder\\
%\footnote{Electronic mail: klauder@phys.ufl.edu}\\
Departments of Physics and Mathematics\\
University of Florida\\
Gainesville, FL  32611}
%Email: klauder@phys.ufl.edu}
\date{}     %   Use   %\date{} to see the dates
\maketitle
\begin{abstract}
At a fixed point in spacetime (say, $x_0$), gravitational phase space 
consists of the space of symmetric matrices $\{F^{ab}\}$ [corresponding 
to the canonical momentum $\pi^{ab}(x_0)$] and of symmetric matrices 
$\{G_{ab}\}$ [corresponding to the canonical metric $g_{ab}(x_0)$], 
where $1 \leq a,b \leq n$, and, crucially, the matrix $\{G_{ab}\}$ is 
necessarily positive definite, i.e. $\sum u^a G_{ab}u^b > 0$ whenever 
$\sum (u^a)^2 > 0$.  In an alternative quantization procedure known as 
{\it Metrical Quantization}, the first and most important ingredient is 
the specification of a suitable metric on classical phase space.  Our 
choice of phase space metrics, guided by a recent study of Affine Quantum 
Gravity, leads to gravitational phase space geometries which possess 
{\it constant scalar curvature} and may be regarded as higher dimensional analogs of the 
Poincar\'{e} plane, which applies when $n=1$.  This result is important 
because phase spaces endowed with such symmetry lead 
naturally via the procedures of Metrical Quantization to acceptable 
Hilbert spaces of high dimension.
\end{abstract}
%\vfill\eject
\section{Introduction}
The canonical phase-space variables of classical gravity have been 
identified as the spatial 
metric with components $g_{ab}(x)\,[= g_{ba}(x)]$ and its canonical 
momentum with components 
$\pi^{ab}(x)\,[=\pi^{ba}(x)]$, where $a,b=1,2,\ldots, n$, in an 
$(n+1)$-dimensional spacetime \cite{adm}. 
While the momentum variables are unrestricted in the values they may 
assume, the metric variables 
are restricted in domain by their physical interpretation so that 
the $(n\times n)$-matrix made 
from the metric components is positive definite for all $x$, i.e., 
$\{g_{ab}(x)\}>0$. Stated 
otherwise, if $u^a$, $a=1,\ldots, n$, denotes an arbitrary vector 
which satisfies $\Sigma\s(u^a)^2>0$, 
then we insist that $u^a\s g_{ab}(x)\s u^b>0$ for all $x$. Therefore, 
while the momentum variables 
$\pi^{ab}(x)$ are elements of a linear vector space, the metric 
variables $g_{ab}(x)$ decidely do 
{\it not} form a linear vector space. These important physical facts 
will have a basic significance 
for our discussion. 

Our use of the word ``gravity'' is strictly motivational and carries 
no implication that the 
metric and momentum variables satisfy either the equations of motion or 
the constraints that 
characterize Einstein's gravitational theory. Instead, our discussion 
applies to a more primitive stage, 
namely, preparing the kinematics of classical phase space to accomodate 
phase-space variables that 
have the domain limitations that characterize the usual gravitational 
variables. Our goal is to 
discuss metrics and their curvature on such phase-space manifolds that 
fully respect these important 
domain restrictions.

Indeed, our motivating interest in phase-space metrics arises from a 
special version of 
quantization. Traditionally, of course, a Riemannian structure on classical 
phase space is not normally 
invoked in quantization, but there is a little-known approach to 
quantization, called {\it Metrical 
Quantization} in which {\it the choice of a Riemannian metric on 
classical phase space is the first 
and key ingredient} in that procedure, and it is that approach which we 
have in mind \cite{kla1}. 

To help the reader get a feel for metrical quantization, let us turn to 
a brief survey of this 
approach for a single degree of freedom system.

\subsection*{Metrical Quantization}
As an illustration of this procedure, we discuss two simple, one degree 
of freedom examples. The 
traditional phase-space path integral, stripped of any specific dynamics, 
formally reads (in units where $\hbar = 1$)
  \bn {\cal M}\int e^{i\tint p\,{\dot q}\,dt}\,\D p\,\D q\;,  \en
where ${\cal M}$ represents a suitable normalization.  This ``integral'' is notoriously ill 
defined.
One way to regularize and thereby give meaning to this formal expression 
is as follows:
  \bn  \<p'',q''|p',q'\>=\lim_{\nu\ra\infty}{\cal N}_\nu\int 
e^{i\tint p\,{\dot 
q}\,dt-(1/2\nu)\tint({\dot p}^2+{\dot q}^2)\,dt}\,\D p\,\D q\;, \label{e2}\en
where now ${\cal N}_\nu$ is a $\nu$-dependent normalization factor.  
In fact, this latter expression can be given a rigorous and unambiguous 
mathematical meaning as
 \bn \<p'',q''|p',q'\>=\lim_{\nu\ra\infty}2\pi\s e^{\nu\s T/2}\,\int 
e^{i\tint 
p\,dq}\,d\mu_W^\nu(p,q)  \label{e3}\en
in terms of a two-dimensional Wiener measure $\mu_W^\nu$ on a flat phase 
space (note, ${\dot 
p}^2+{\dot q}^2\propto dp^2+dq^2$), where the parameter $\nu$ denotes the 
diffusion constant. The 
result of this path integral over phase-space paths $p(t),\, q(t)$, 
$0<t<T$, is a kernel, which we have 
already called $\<p'',q''|p',q'\>$, and which depends on the initial and 
final pinned values for 
each of the two paths, i.e., $p(0)=p',\, q(0)=q'$ and $p(T)=p'',\,q(T)=q''$. 
As the notation is 
intended to suggest, the result of the path integral is in fact just the 
overlap of two {\it 
canonical coherent states} of the form
  \bn  |p,q\>\equiv e^{-iq\s P}\s e^{ip\s Q}\,|0\>  \;,  \label{e4}\en
where the fiducial vector $|0\>$ is a unit vector that satisfies 
$(Q+iP)\,|0\>=0$. As usual, $Q$ 
and $P$ denote self-adjoint Heisenberg operators that satisfy $[Q,P]=i\one$. 
It is important to 
understand that all the properties implicit in (\ref{e4}) are a 
{\it direct consequence} of the 
assumed form of the regularization---specifically, the form of the 
{\it phase-space 
metric}---introduced into (\ref{e2}). This fact follows because the 
metric determines the kernel, and through the GNS (Gel'fand, Naimark, 
and Segal) Theorem \cite{emch}, the kernel determines the form of 
(\ref{e4}) up to unitary equivalence. Thus, 
the very choice of the metric, specific to a flat space expressed in 
Cartesian coordinates, has, 
through (\ref{e2}) and (\ref{e3}), led to an explicit quantization in 
terms of canonical Heisenberg 
variables $P$ and $Q$ as the basic kinematical operators. 

It is noteworthy that a change of the metric from $d\sigma^2=dp^2+dq^2$ 
to $d\sigma^2=\lambda^{-1}\s 
q^2\,dp^2+\lambda\s q^{-2}\s dq^2$, $q>0$, $\lambda>\half$, leads to an 
{\it entirely different 
quantum kinematical structure}.
We first observe that the geometry of the phase-space manifold under the 
second metric is a 
simply-connected space of constant scalar curvature $-2\lambda^{-1}$ (the {\it 
Poincar\'{e} plane}), and which, therefore, constitutes a 
distinctly different geometry than that of the flat case. If we use the 
new metric to support the 
Brownian motion paths, then the corresponding phase-space path integral 
reads
\bn \<p'',q''|p',q'\>&=&\lim_{\nu\ra\infty}{\tilde{\cal N}}_\nu\int 
e^{i\tint p\,{\dot 
q}\,dt-(1/2\nu)\tint(\lambda^{-1}\s q^2\,{\dot p}^2+\lambda\s q^{-2}\s 
{\dot q}^2)\,dt}\,\D p\,\D q\no \\
&=&\lim_{\nu\ra\infty}2\pi[1-(2\lambda)^{-1}]^{-1}\s e^{\nu\s T/2}\,
\int e^{i\tint 
p\,dq}\,d{\tilde\mu}_W^\nu(p,q) 
\label{e5}\;.  \en
Here ${\tilde\mu}_W^\nu$ is also a Wiener measure on a two-dimensional 
manifold but now a manifold 
of constant scalar curvature. Observe that the second form of (\ref{e5}) has a 
rigorous and 
unambiguous meaning. The remarkable fact, thanks again to the GNS Theorem, 
is that the phase-space 
path integral (\ref{e5}) determines a kernel that is given this time by the 
overlap of {\it affine 
coherent states} as defined by the expression
  \bn  |p,q\>\equiv e^{ip(Q-q)}\s e^{-i\ln(q)\s D}\,|0\>\;,\hskip1cm q>0\;, 
\en
where $[Q,D]=iQ$, $Q>0$, and $|0\>$ is a normalized solution of 
$[D-i\lambda(Q-1)]\s|0\>=0$. Thus 
the second metric choice has resulted in an explicit quantization in 
terms of affine Lie algebra 
variables $D$ and $Q$ as the basic kinematical operators (rather than 
the more common $P$ and $Q$).

Finally, let us observe that it is possible to consider the expression
 \bn \lim_{\nu\ra\infty}{\cal N}'_\nu\int e^{i\tint p\s{\dot 
q}\,dt-(1/2\nu)\tint[d\sigma(p,q)^2/dt^2]\,dt}\,\D p\,\D q\;, 
\label{e12}\en
for a general Riemannian metric
\bn d\sigma(p,q)^2\equiv A(p,q)\, dp^2+2\s B(p,q)\s dp\s dq+C(p,q)
\s dq^2\;, \en
where $A>0$, $C>0$, and $AC>B^2$, and to ask what quantum Hilbert space 
might then apply. Although we have couched the question in 
terms of a phase-space path integral, by the Feynman-Kac Theorem we can 
reformulate the question in 
terms of a second order partial differential equation. For example, the 
path integral (\ref{e12}) 
may be interpreted in terms of a two-dimensional particle whose kinetic 
energy is 
$(1/2\nu)\tint[d\sigma(p,q)^2/dt^2]\,dt$ moving in the presence of a 
uniform magnetic field. The dimensionality of 
the quantum Hilbert space is the same dimensionality as the degeneracy 
of the lowest level (ground 
state) of the associated Hamiltonian. 

\subsection*{Generalized Affine Coherent States and Gravitational Phase 
Space}

Motivated by the desire for a quantum kinematical framework in which the 
aforementioned algebraic restrictions on the spacial portion of the 
metric of general relativity are automatically respected 
\cite{affinegravity},  we have recently examined an interesting matrix 
generalization of the affine algebra in which the spectrum of any line 
element is {\it manifestly positive}.  This algebra generates a family 
of coherent states ({\it generalized affine coherent states}) whose 
overlap function may be represented as [cf. (5)]
\bn \nonumber \<F'', G''|F', G'\>&\\ 
\nonumber && \hskip -3cm = \lim_{\nu \rightarrow \infty} {\cal M_{\nu}} 
\int e^{i\tint \{-{\rm tr}(G {\dot F}) \} dt \; - \; (1/2\nu) 
\tint \{ \lambda^{-1}{\rm tr}[(G {\dot F})^2] \; + \; \lambda \; 
{\rm tr}[(G^{-1}{\dot G})^2] \} \; dt} \\
&& \hskip +3cm \times\prod_{j\leq k} {\cal D}F^{jk}\;                 
{\cal D}G_{jk},
\en
where now $F \equiv \{F^{jk}\}$ is an $n\times n$ symmetric matrix, 
$G \equiv \{G_{jk}\}$ is an $n\times n$, {\it positive}, symmetric matrix, 
and $\lambda$ is a scaling parameter; for the precise meaning of 
(\theequation), as well as a discussion of related matters, see 
\cite{wat}.  The phase space metric appearing in the regularization 
factor in (\theequation) is described by  

\bn \nonumber d\sigma^2 &=& \lambda^{-1}{\rm tr}[(G \s dF)^2] + \lambda 
\; {\rm tr}[(G^{-1}\s dG)^2]\\
&=& \lambda^{-1} G_{kl}\s G_{mj} dF^{jk}\s dF^{lm}  + \lambda \; G^{kl}\s G^{mj} 
dG_{jk}\s dG_{lm} ,
\en
where the indices $j$, $k$, $l$, $m$, take on values from $1$ to $n$.\footnote {It is also of 
interest to note that the metric in (10) is formally identical to 
\bn \nonumber
||d|F,G\>||^2 - |\<F,G|d|F,G\>|^2,
\en
the infinitesimal ray metric on the corresponding Hilbert Space.  This relation makes clear that 
$d\sigma^2$ is invariant under all right translations induced by the affine group that defines the 
coherent states in \cite{wat}, and therefore characterizes a homogeneous space.}  
The nonzero components of the phase space metric may 
thus be written explicity as
\bn \nonumber g_{F^{jk}F^{lm}} &=& \half\s\lambda^{-1}\s(G_{kl}\s G_{mj}+G_{jl}\s G_{mk}),\\
g_{G_{jk}G_{lm}} &=& \half\s\lambda\s(G^{kl}G^{mj}+G^{jl}\s G^{mk}),
\en
and its determinant is a function only of the dimensionality, namely,
\bn
\det g = 2^{n(n-1)}.  
\en
We shall refer to the $n(n+1)$-dimensional manifold spanned by the 
coordinates $F^{jk} (j \leq k)$ and $G_{lm} (l \leq m)$, with 
$\{G_{lm}\} > 0$, as {\it gravitational phase space}.  In what follows, 
we shall also implicitly include a metric on gravitational phase space as 
part of its definition.  

{}From the discussion in the previous section it should be clear that the 
geometry of gravitational phase space is of crucial importance in 
determining the 
nature of the associated Hilbert space as determined by (9).  We now 
turn to an investigation of this geometry.   

\section{Curvature Calculations}

In this section we outline the calculation leading to the scalar 
curvature of a gravitational phase space of general dimensionality whose 
metric is described by (10).  It first proves convenient, however, not to 
enforce the customary symmetry conditions $F^{kj} = F^{jk}$ and 
$G_{ml} = G_{lm}$, thus constructing a $2n^2$-dimensional manifold 
which we shall refer to as the {\it extended phase space}.  We show that 
the scalar curvature\footnote{For convenience, we follow a sign convention 
for the scalar curvature in which the scalar curvature of the 
Poincar\'{e} plane is {\it negative}.  See \cite{Weinberg} for a full 
discussion.}
\bn R_E = -2n^3 \lambda^{-1}.
\en

The corresponding result for the space of symmetric matrices (where we again enforce 
$F^{kj} = F^{jk}$ and $G_{ml} = G_{lm}$), may be extracted from the 
nonsymmetric calculation via a careful process of symmetrization, 
whereby all the terms associated with skew-symmetric directions on the 
manifold are eliminated.  We thus show that the scalar curvature of 
gravitational phase space is 
\bn R = -\half n(n+1)^2 \lambda^{-1}.
\en
The correctness of the results in (13) and (\theequation) may be verified 
computationally for at least the first few values of $n$.

\subsection*{Curvature of the Extended Phase Space}
We begin by discussing the geometry of the $2n^2$-dimensional manifold whose 
coordinates may be taken to be the $n^2$ elements of the matrix $F$ together 
with the $n^2$ elements of the matrix $G$.  At this point, we treat 
$F^{jk}$ and $F^{kj}$ (and likewise $G_{lm}$ and $G_{ml}$) as completely 
independent coordinates - the restriction to the submanifold of symmetric matrices is descibed in 
the following subsection.

The connection coefficients associated with the metric in (10) may be 
calculated by using the standard trick of considering the dynamics of a 
classical free particle moving in the corresponding $2n^2$-dimensional 
geometry.  The motion of such a particle is governed by a Lagrangian of 
the form
\bn L = \half{\rm tr}[\lambda^{-1}(G \dot{F})^2 + \lambda (G^{-1} 
\dot{G})^2],
\en  
leading to equations of motion for $F$ and $G$ given by
\bn\nonumber \ddot{F} &=& -(\dot{F}\s \dot{G}\s G^{-1} + G^{-1}\s\dot{G}\s\dot{F}),\\
\ddot{G} &=& \lambda^{-2} G\s\dot{F}\s G\s\dot{F}\s G + 
\dot{G}\s G^{-1}\s\dot{G}.
\en
A re-interpretation of (\theequation) as geodesic equations quickly 
reveals the form of the nonzero symmetric connection coefficients as
\bn \Gamma^{G_{jk}}_{F^{ab}F^{cd}} &=& \Gamma^{G_{jk}}_{F^{cd}F^{ab}} 
= -\half \lambda^{-2}(G_{ja}G_{bc}G_{dk} + G_{jc}G_{da}G_{bk}),\\
\Gamma^{F^{jk}}_{G_{ab}F^{cd}} &=& \Gamma^{F^{jk}}_{F^{cd}G_{ab}} = 
\half (\delta_d^k \delta_c^b G^{ja} + \delta_c^j \delta_d^a G^{bk}),\\
\Gamma^{G_{jk}}_{G_{ab}G_{cd}} &=& \Gamma^{G_{jk}}_{G_{cd}G_{ab}} = 
-\half(G^{bc}\delta_j^a \delta_k^d + G^{da} \delta_j^c \delta_k^b).
\en
Ricci tensor elements may be constructed from the connection coefficients 
via the well-known relation \cite{Schutz}
\bn R_{\alpha \beta} = \partial_{\gamma} \Gamma_{\alpha \beta}^{\gamma} 
- \partial_{\beta} \Gamma_{\gamma \alpha}^{\gamma} + 
\Gamma_{\gamma \delta}^{\gamma} \Gamma_{\alpha \beta}^{\delta} - 
\Gamma_{\beta \gamma}^{\delta} \Gamma_{\delta \alpha}^{\gamma}.
\en
The second and third terms on the right hand side of (\theequation) 
vanish as a consequence of the constant nature of the determinant of 
the metric in (10).  Evaluation of the remaining terms yields
\bn
\nonumber R_{F^{jk}F^{lm}} &=&  \frac{\partial}{\partial G_{ab}} 
\Gamma^{G_{ab}}_{F^{jk}F^{lm}} - \Gamma^{G_{ab}}_{F^{cd}F^{jk}}
\Gamma^{F^{cd}}_{F^{lm}G_{ab}} - \Gamma^{F^{ab}}_{G_{cd}F^{jk}} 
\Gamma^{G_{cd}}_{F^{lm}F^{ab}}\\
&=& -n \lambda^{-2} G_{mj}G_{kl},\\
\nonumber R_{G_{jk}G_{lm}} &=& \frac{\partial}{\partial G_{ab}} 
\Gamma^{G_{ab}}_{G_{jk}G_{lm}} - \Gamma^{G_{ab}}_{G_{cd}G_{jk}} 
\Gamma^{G_{cd}}_{G_{lm}G_{ab}} - \Gamma^{F^{ab}}_{F^{cd}G_{jk}} 
\Gamma^{F^{cd}}_{G_{lm}F^{ab}} \\
&=& -nG^{mj}G^{kl}.
\en
Finally, contraction with the inverse metric then yields the 
scalar curvature of the extended phase space manifold,
\bn \nonumber R_E &=& g^{F^{jk}F^{lm}}R_{F^{jk}F^{lm}} + 
g^{G_{jk}G_{lm}}R_{G_{jk}G_{lm}}\\
\nonumber &=& \lambda G^{jm}G^{lk}R_{F^{jk}F^{lm}} + \lambda^{-1} 
G_{jm}G_{lk}R_{G_{jk}G_{lm}}\\
\nonumber &=& (-n^3 \lambda^{-1}) + (-n^3 \lambda^{-1})\\
&=& -2n^3 \lambda^{-1}.
\en

It is worth pointing out that the $F$-$G$ crossterms of (18) play a 
vital role in the calculation above - the scalar curvature of the 
corresponding $n^2$-dimensional manifold involving only the elements 
of the matrix $G$ (and not those of $F$) turns out to be 
$-\half n(n^2-1) \lambda^{-1}$, not simply $-n^3 \lambda^{-1}$, 
as a cursory glance at (\theequation) might suggest.

\subsection*{Curvature of the Phase Space of Symmetric Matrices}
We now restrict attention to the submanifold of symmetric matrices defined by the 
restrictions $F^{kj} \equiv F^{jk}$ and $G_{ml} \equiv G_{lm}$.  The 
metric of (10) again applies, subject of course to the stipulations 
that the variations $dF$ and $dG$ should likewise satisfy  
$dF^{kj} \equiv dF^{jk}$ and $dG_{ml} \equiv dG_{lm}$.

The curvature of the submanifold of symmetric matrices may again be obtained starting 
from (20).  This time however it is necessary to employ {\it symmetric 
contractions} in (21) and (22) in order to filter out those contributions 
present in the previous calculation involving skew-symmetric directions.  
This process yields the Ricci tensor elements
\bn R_{F^{(jk)}F^{(lm)}} &=& -\half(n+1) \lambda^{-2} G_{(m(j}G_{k)l)}\\ 
R_{G_{(jk)}G_{(lm)}} &=& -\half(n+1)G^{(m(j}G^{k)l)}
\en
and scalar curvature
\bn \nonumber R &=& g^{F^{(jk)}F^{(lm)}}R_{F^{(jk)}F^{(lm)}} + 
g^{G_{(jk)}G_{(lm)}}R_{G_{(jk)}G_{(lm)}}\\
\nonumber &=& [-n(n+1)^2 \lambda^{-1}/4] + [-n(n+1)^2 \lambda^{-1}/4]\\
&=& -\half n(n+1)^2 \lambda^{-1}.
\en

The two-dimensional result, $R(n=1) = R_E(n=1) = -2\lambda^{-1}$, 
will be recognized as the constant negative scalar curvature of the 
Poincar\'e plane.  Notice that the scalar curvature of the generalized 
higher-dimensional phase space manifolds we have described remains 
constant for all n.

We emphasize that $F$-$G$ crossterms of (18) are again important in the 
derivation of (\theequation); in particular, the scalar curvature of the 
corresponding submanifold involving only the symmetric matrix 
$G$ is $-n(n-1)(n+2)/8\lambda$.

\section{Conclusion}

The phase space metrics described by (10) define gravitational phase 
spaces of constant scalar curvature, the key ingredient necessary to induce, via (9), the infinite 
dimensional Hilbert spaces associated with the generalized affine algebra discussed in \cite{wat}.

The physical concept behind our continuous-time regularization procedure involves the motion of a 
free particle on a curved space (in which our phase space is regarded as the configuration space).  
Quantization of such a system leads to the Laplace-Beltrami operator plus a possible additional 
term of order $\hbar^2$ proportional to the {\it scalar curvature} (see, e.g., \cite{maraner}).  As 
our scalar curvature is constant, this term represents a harmless factor in the integral (9) that 
can be included in the overall normalization.  Higher symmetry, such as that represented by 
constant {\it sectional} curvature, is not required to obtain this result.  In fact, although our 
metrics (10) do indeed possess the high degree of symmetry inherited from their group definition, they 
do not (for $n\geq2$) define spaces of constant sectional curvature, since the identification of 
the various coordinates as elements of a positive definite matrix implies a lack of isotropy on the 
manifold.  

The type of argument we have presented can also be used to disqualify a large class of candidate 
phase space metrics from serious 
consideration.  For example, a few minutes with a tensor manipulation 
program will be enough to convince the reader that the plausible-looking 
phase space metric \cite{plausible} described by 
\bn
 d\sigma^2 &=& (\det G)^{-1/2} \s 
{\rm tr}[(G \s dF)^2] + (\det G)^{1/2} \s {\rm tr}[(G^{-1}\s dG)^2]
\en
does {\it not} lead to a gravitational phase space of constant scalar curvature, 
and therefore cannot generate via an analog of (9) anything other than a 
Hilbert space of trivial dimensionality. 

\section*{Acknowledgements}
Partial support for this work from NSF Grant PHY-0070650 is gratefully 
acknowledged.  It is a pleasure to express thanks to David Metzler for valuable discussions and 
also to the referees for important insight and suggestions.

\end{document}